\documentclass[12pt,epsf,epsfig,aps,floatfix,preprint,nofootinbib]{revtex4-1}
\usepackage{graphics}
\usepackage{graphicx}
\usepackage[english]{babel}
\usepackage{fullpage}
\usepackage{amssymb}
\usepackage{amsmath}
\usepackage{verbatim}
\usepackage{pstricks}
\usepackage{slashed}

\newcommand{\nc}{\newcommand}
\nc{\beq}{\begin{equation}}
\newcommand{\mx}{m_{X}}

\newcommand{\vesc}{v_{esc}}
\newcommand{\vb}{\bar{v}}

\begin{document}

\setlength{\baselineskip}{0.22in}

\begin{flushright}MCTP-11-16\\
\end{flushright}

\vspace{0.2cm}

\title{Constraints on Scalar Asymmetric Dark Matter from Black Hole Formation in Neutron Stars}

\author{
Samuel D. McDermott, Hai-Bo Yu, Kathryn M. Zurek
}

\vspace*{0.2cm}

\affiliation{
Michigan Center for Theoretical Physics, Department of Physics, University of Michigan, Ann Arbor, Michigan 48109, USA
}

\date{\today}

\begin{abstract}
\noindent
We consider possibly observable effects of asymmetric dark matter (ADM) in neutron stars.  Since dark matter does not self-annihilate in the ADM scenario, dark matter accumulates in neutron stars, eventually reaching the Chandrasekhar limit and forming a black hole.  We focus on the case of scalar ADM, where the constraints from Bose-Einstein condensation and subsequent black hole formation are most severe due to the absence of Fermi degeneracy pressure.  We also note that in some portions of this constrained parameter space, nontrivial effects from Hawking radiation can modify our limits. We find that for scalar ADM with mass between 5 MeV and 13 GeV, the constraint from nearby neutron stars on the scattering cross section with neutrons ranges from $\sigma_n \lesssim 10^{-45} \mbox{ cm}^2$ to $10^{-47} \mbox{cm}^2$.


\end{abstract}

\maketitle

\section{Introduction}

The characteristics of dark matter (DM) and the nature of its production mechanism have so far eluded description. While cosmological observations provide compelling evidence for its existence, its mass and the nature of its interactions with the standard model remain unknown. One popular hypothesis holds that the DM interacts with standard model particles via the weak interaction and is also self-annihilating. In this weakly interacting massive particle (WIMP) scenario~\cite{review}, the correct relic density of DM is a natural consequence of the thermal history of the early Universe.

Alternatively, DM may carry a conserved charge, analogous to baryon number. This asymmetric DM (ADM) scenario is motivated by the fact that the DM and baryon densities are of the same order of magnitude. The earliest models attempting to relate the DM to the baryon asymmetry made use of electroweak sphalerons \cite{Early} or out-of-equilibrium decay \cite{outofequilibrium}.  The former often run into tight constraints from LEP measurements.  In contrast, the modern incarnation of ADM makes use of higher dimension operators to transfer the asymmetry in a robust way that is relatively free of electroweak constraints \cite{Asymmetric}.  ADM models prefer to have DM mass around a few GeV (see for example~\cite{models,Cohen:2010kn,Kang:2011wb}), which is consistent with hints from recent direct detection experiments~\cite{Bernabei:2008yi,Aalseth:2010vx}.  It is also possible for ADM to have weak scale mass \cite{quirky,Lepto}, or mass well below a GeV \cite{Falkowski:2011xh}.

The DM mass and its scattering cross section with nuclei have been constrained by various underground direct detection experiments~\cite{Ahmed:2009zw,Angle:2007uj,Lebedenko:2008gb}, as well as by particle colliders~\cite{Beltran:2008xg,Cao:2009uw,Goodman:2010yf,Bai:2010hh}.  In this article, we study the properties of ADM through its impacts on stellar systems.  It has long been appreciated that a finite DM-nucleon cross section would result in DM capture in stars~\cite{Press:1985ug,Gould:1987ju,Griest:1986yu}. In the WIMP scenario, DM annihilation can generate an additional heat source, which may affect stellar formation~\cite{Spolyar:2007qv} and evolution \cite{Iocco:2008rb}, or cause anomalous heating of white dwarfs~\cite{McCullough:2010ai,Hooper:2010es} and neutron stars~\cite{Kouvaris:2007ay,Kouvaris:2010vv,deLavallaz:2010wp}. In the ADM case, DM particles do not annihilate and hence provide no additional power for stars. However, since there is no annihilation to deplete ADM particles, stars can accumulate far more ADM particles than usual WIMPs, which can lead to different effects. For example, it could have implications for solar physics~\cite{Frandsen:2010yj,Cumberbatch:2010hh,Taoso:2010tg}, or change the mass-radius relation of neutron stars~\cite{Sandin:2008db,Ciarcelluti:2010ji,Iorio:2010hb}.  The most extreme possibility is that captured particles can become self-gravitating, forming a black hole that will eventually destroy the host stars~\cite{Goldman:1989nd,Gould:1989gw,Bertone:2007ae}.

Recently, constraints on fermionic ADM through the survival of compact stars have been discussed in~\cite{deLavallaz:2010wp,Kouvaris:2010jy}. In a certain class of ADM models, the DM candidate is a boson~\cite{Cohen:2010kn,Kang:2011wb}. In this paper, we study constraints on this scalar ADM from compact stars. Scalar DM particles differ from fermions by spin statistics, which has a significant impact on black hole formation conditions.  Black hole formation occurs only when the total number of self-gravitating DM particles is larger than the Chandrasekhar limit~\cite{Shapiro:1983du}.  Fermions obey the Pauli exclusion principle, and the Chandrasekhar limit is set by the balance between gravity and the Fermi pressure, while scalar particles have no Fermi pressure to hinder gravity. In this case, the lower limit for gravitational collapse is determined by the balance between gravity and the pressure induced by the zero point energy, which is much smaller than the Pauli pressure experienced by fermions.  Therefore, we derive much stronger constraints on scalar ADM from compact stars. Since neutron stars have much higher matter density and escape velocity than any other stars, we will mainly focus on neutron stars. Note that neutron star constraints on scalar DM has been discussed in~\cite{Goldman:1989nd}. In this early work, the cross section for the DM capture is set by the geometric cross section, and the DM mass range is between $1~{\rm GeV}$ and $100~{\rm TeV}$. In this work, we treat the DM-neutron cross section as a free parameter and use neutron stars to constrain it.  Other important considerations that we treat here are the effect of Bose-Einstein condensation, which significantly alters the constraint derived on the DM-neutron scattering cross-section, and Hawking radiation, which modifies our constraints for high mass DM. We also explore a wider DM mass range and use the most recent neutron star data.

This paper is organized as follows. In Sec. II, we discuss the Chandrasekhar limit for fermions and bosons. In Sec. III, we discuss ADM capture in neutron stars. In Sec. IV, we discuss thermalization, condensation and black hole formation of captured scalar ADM in neutron stars. In Sec. V, we discuss Hawking radiation and destruction of the host neutron star. In Sec. VI, we discuss observational constraints. We present our conclusions in Sec. VII.

\section{Chandrasekhar Limit}

First, we review the derivation of the Chandrasekhar limit for a system of fermions. Suppose there are $N$ fermions of mass $m$ distributed in a sphere with radius $R$, so that the number density of fermions is approximately $N/R^3$.  Because of the Pauli exclusion principle, the average distance between two fermions is $\sim R/N^{1/3}$.  The uncertainty principle requires that each fermion have Fermi momentum $p\sim N^{1/3}/R$. If the total number $N$ is small and $m>p\sim N^{1/3}/R$, the system is in the nonrelativistic limit. The average energy per fermion is
\begin{equation}
E\sim-\frac{GNm^2}{R}+\frac{1}{m}\left(\frac{N^{1/3}}{R}\right)^2,
\end{equation}
where $G$ is Newton's constant. Once the gravitational and Fermi pressures reach equilibrium, the system can have a stable spherical configuration with radius
\begin{equation}
R\sim\frac{1}{G m^3 N^{1/3}}.
\label{radius}
\end{equation}
As $N$ increases, the radius shrinks and the Fermi momentum increases; eventually fermions become relativistic with total energy
\begin{equation}
E\sim-\frac{GNm^2}{R}+\frac{N^{1/3}}{R}.
\label{fermionenergy}
\end{equation}
If the total number of the fermions increases beyond the limit
\begin{equation}
N^{fermion}_{Cha}\sim\left(\frac{1}{Gm^2}\right)^{3/2}=\left(\frac{M_{pl}}{m}\right)^3\simeq1.8\times10^{51}\left(\frac{100~{\rm GeV}}{m}\right)^{3},
\label{fermionlimit}
\end{equation}
where $M_{pl}=1.2211\times10^{19}~{\rm GeV}$ is the Planck scale, the gravitational energy will dominate the total particle energy and gravitational collapse will occur. This is the famous Chandrasekhar limit~\cite{Shapiro:1983du}.

Now we discuss bosons. Similar to the fermion case, the gravitational collapse occurs when particles are relativistic. But the bosonic system is significantly different from the fermionic system because it has no Fermi pressure to hinder gravity. Since the bosons are confined inside a sphere with radius $R$, they have zero point energy $1/R$ due to the uncertainty principle in the relativistic limit. Therefore, the typical energy for a boson in a sphere of radius $R$ is
\begin{equation}
E\sim-\frac{GNm^2}{R}+\frac{1}{R}.
\end{equation}
Again, the radius cancels in the critical limit.
In this case, the Chandrasekhar limit is
\begin{equation}
\label{bosonlimit}
N^{boson}_{Cha}\simeq \left(\frac{M_{pl}}{m}\right)^{2}\simeq1.5\times10^{34}\left(\frac{100~{\rm GeV}}{m}\right)^{2}.
\end{equation}
Comparing Eq. (\ref{fermionlimit}) and Eq. (\ref{bosonlimit}), we can see that for a given particle mass, a particle that obeys Bose-Einstein statistics will experience gravitational collapse much more readily than a particle that obeys Fermi-Dirac statistics.

When the total number of DM particles accumulated in a neutron star surpasses the Chandrasekhar limit, the captured DM particles collapse to a black hole and destroy the host neutron star. Therefore, observations of old neutron stars can be used to constrain the DM-neutron scattering cross section. Since bosons have much smaller Chandrasekhar limit than fermions, we can obtain stronger limits on bosonic DM. In this work, we take typical neutron star parameters $M_n=1.44~M_{\odot}$, $R_n=10.6~{\rm km}$ and the central density $\rho_B=1.4\times10^{15}~{\rm g/cm^{3}}$~\cite{deLavallaz:2010wp,Clark:2002db}.

\section{Capture of Asymmetric Dark Matter in Neutron Stars}

The accretion of DM onto stars has been studied in~\cite{Press:1985ug,Gould:1987ju,Griest:1986yu}. In this section, we review the basic formulas for the capture of asymmetric DM in neutron stars.

In the absence of annihilation, the number of DM particles in a star is determined by the differential equation
\begin{equation}
\frac{dN_X}{dt}=C_B,
\label{diffeq}
\end{equation}
where $N_X$ is the total number of DM particles in the star and $C_B$ is the DM capture rate through scattering with baryons. In this study, we assume there is no symmetric component and neglect DM annihilation completely. In some ADM models, a portion of the symmetric component can be regenerated by oscillation effects \cite{Lepto,Falkowski:2011xh}. For those models, bounds derived from the survival of the neutron stars can be weaker.

Additionally, we ignore the self-capture effect. Dark matter may have a sizable self-interaction that leads to self-capture~\cite{Zentner:2009is}. However, the self-capture saturates when the sum of the individual self-scattering cross sections becomes larger than the geometrical area over which the DM particles thermally distribute. As we will show in the next section, due to the large baryon density the captured DM particles are thermally distributed in a very small region of radius $\sim 1~{\rm m}$ in the core of the neutron star. We have checked that the baryonic capture always dominates the DM accretion process in neutron stars for the parameter space of interest, even if we take the upper limit of the self-scattering cross section allowed by the elliptical halo shape bound~\cite{Feng:2009hw}.

\subsection{Capture rate}
The number of DM particles in the neutron star is set by $C_B$. Since neutrons are degenerate in the neutron stars, capture can occur only when the momentum transfer is larger than the difference between the Fermi momentum and the neutron momentum. As we will show, this will affect the capture efficiency significantly for DM with mass less than $1~{\rm GeV}$. For larger DM mass the effect is negligible, because the momentum transfer is always sizable.

The accretion rate $C_B$ is given by \cite{Gould:1987ju}
\begin{equation}
C_B=4\pi\int^{R_{n}}_0r^2\frac{dC_B(r)}{dV}dr,
\end{equation}
where $R_{n}$ is the radius of the neutron star and the capture rate per unit volume for an observer at rest with respect to the DM distribution is given by
\begin{equation}
\frac{dC_B(r)}{dV}=\sqrt{\frac{6}{\pi}}n_X(r)n_B(r)\xi\frac{v(r)^2}{\vb^2}(\vb\sigma_{XB})\left[1-\frac{1-\exp\left(-B^2\right)}{B^2}\right].
\end{equation}
Here $n_X(r)$ is the ambient DM number density; $n_B(r)$ is the number density of the stellar baryons; $\vb$ is the DM velocity dispersion around the neutron star; $v(r)$ is the escape velocity of the neutron star at the given radius $r$; $\sigma_{XB}$ is the effective scattering cross section between DM particles and nucleons in the neutron star; and $\xi$ takes into account the neutron degeneracy effect on the capture. The factor $B^2$ is given by
\begin{equation}
B^2=\frac{3}{2}\frac{v(r)^2}{\vb^2}\frac{\mu}{\mu^2_-},
\end{equation}
where $\mu=\mx/m_B$ and $\mu_-=(\mu-1)/2$.

Now we specify the factor $\xi$. All energy levels below the Fermi momentum $p_F$ have been occupied. During the scattering process, if the momentum transfer to the neutron is larger than $p_F$, the scattered neutron can be excited above the Fermi surface. In this case, all neutrons can participate the capture process, and the capture efficiency  is $\xi=1$. On the other hand, if the momentum transfer $\delta p$ is less than $p_F$, only those neutrons with momentum larger than $\sim p_F-\delta p$ can participate in the capture process. The fraction of these neutrons is $\sim\delta p/p_F$, so we can approximate $\xi$ as $\xi\simeq \delta p/p_F$. Depending upon the momentum transfer $\delta p$, we can parameterize $\xi$ as
\begin{equation}
\label{saturation}
\xi={\rm Min}\left[\frac{\delta p}{p_F},1\right].
\end{equation}
When the DM particle approaches the surface of the neutron star, its velocity is close to the escape velocity. Hence, the typical momentum transfer is $\delta p\simeq \sqrt{2}m_rv_{esc}$, where $m_r=\mx m_B/(\mx+m_B)$ is the reduced mass, and typically $\vesc \simeq 1.8\times10^{5}~{\rm km/s}$. The Fermi momentum is $p_F\simeq(3\pi^2\rho_B/m_B)^{1/3}\simeq0.575~{\rm GeV}$ for $\rho_B=1.4\times10^{15}~{\rm g/cm^{3}}$. Therefore, $\xi \simeq1$ for all $\mx\gtrsim 1~{\rm GeV}$. In contrast, the capture rate is suppressed by a factor $\sim\mx v_{esc}/p_F$ if the DM mass smaller than the neutron mass.

To estimate the capture rate, we take the conservative limit that $v(r)=v(R_n) \equiv \vesc$, and we assume that $n_X(r)$ and $n_B(r)$ are independent of the radius; thus, the total capture rate can be simplified to
\begin{equation}
C_B\simeq\sqrt{\frac{6}{\pi}}\frac{\rho_X}{\mx}\frac{\vesc^2}{\vb^2}(\vb\sigma_{XB})\xi N_B\left[1-\frac{1-\exp\left(-B^2\right)}{B^2}\right],
\label{rate}
\end{equation}
where $N_B$ is the total number of neutrons in the host star. When $B^2\gg1$, the term in the square bracket is close to 1; for typical values $\vesc \simeq 1.8\times10^{5}~{\rm km/s}$ and $ \vb \simeq 220~{\rm km/s}$, and a DM mass smaller than $9.4\times10^{5}~{\rm GeV}$, this condition is obtained. If DM has mass larger than $\sim9.4\times10^{5}~{\rm GeV}$, a lower probability to lose enough kinetic energy to be captured after single scatter results~\cite{deLavallaz:2010wp}. In our numerical work, we use the full expression of Eq.~(\ref{rate}).

\subsection{Total Number of ADM in Neutron Stars}

Here we only consider DM particles captured by the neutron star itself and neglect those the neutron star can inherit from its progenitor. Compared to the neutron star phase, the progenitor usually has much lower density and shorter lifetime which results in lower capture efficiency. The total number of DM particles captured by the neutron star is given by the solution of Eq.~\eqref{diffeq}
\begin{equation}
N_X=C_B t.
\end{equation}
To evaluate $C_B$, we note that if the sum of individual nucleon-DM scattering cross sections is larger than the geometric surface area of the star, the capture rate will saturate. Therefore, the capture rate increases with the cumulative nucleon-DM scattering cross section $\sigma_{tot}=N_B\sigma_n$, where $\sigma_n$ is the DM-neutron elastic scattering cross section, as long as $\sigma_{tot}$ is smaller than $\sigma_{geom}=\pi R^2_{n}$; that is, we can constrain the individual scattering cross section $\sigma_n$ as long as $\sigma_n$ is less than or equal to $\sigma_{max}=\pi R^2_{n}/N_B$. Taking typical neutron star parameters $M_n=1.44~M_{\odot}$ and $R_n=10.6~{\rm km}$, we estimate the maximum cross section as~\cite{deLavallaz:2010wp}
\begin{equation}
\sigma_{max}=2.1\times10^{-45}~{\rm cm^2}\left(\frac{R_n}{10.6~{\rm km}}\right)^2\left(\frac{1.44~M_\odot}{M_n}\right),
\end{equation}
and the effective cross section is given by
\begin{equation}
\sigma_{XB}={\rm Min}\left[\sigma_n,\sigma_{max}\right].
\end{equation}
Note that since we consider scattering off only one nucleon, this scattering can be regarded as either spin-dependent or spin-independent.

Now we can estimate the total number of ADM in the neutron star at a given time, using generic parameters $\vesc=1.8\times10^5~{\rm km/s}$, $\bar{v}=220~{\rm km/s}$, and $N_B\simeq1.7\times10^{57}$. In the regime $\mx\gtrsim1~{\rm GeV}$, we have $\xi\simeq1$, which gives
\begin{equation}
N_X\simeq2.3\times10^{44}\left(\frac{100~{\rm GeV}}{\mx}\right)\left(\frac{\rho_X}{10^3~{\rm GeV/cm^3}}\right)\left(\frac{\sigma_{XB}}{2.1\times10^{-45}~{\rm cm^2}}\right)\left(\frac{t}{10^{10}~{\rm years}}\right).
\label{NXlarge}
\end{equation}
When the DM mass is less than $\sim1~{\rm GeV}$, the degeneracy effect on the capture process is important so that $\xi\simeq\sqrt{2}\mx v_{esc}/p_F$, and we have
\begin{equation}
N_X\simeq3.4\times10^{46}\left(\frac{\rho_X}{10^3~{\rm GeV/cm^3}}\right)\left(\frac{\sigma_{XB}}{2.1\times10^{-45}~{\rm cm^2}}\right)\left(\frac{t}{10^{10}~{\rm years}}\right).
\label{NXsmall}
\end{equation}
It is interesting to note that the DM number does not depend on the DM mass in the second case.

In the above derivation of $N_X$, we have assumed that the evaporation effect is negligible for the DM. Now we estimate the DM mass scale below which the evaporation is relevant. Since energy states below the Fermi surface are occupied, only those neutrons with momentum above $p_F$ can transfer kinetic energy to the DM. Since $T\ll p_F$ for the neutron star, the number of these free neutrons is order $\sim10^{-8}$ smaller than that of the neutrons in the Fermi sea. So the scattering probability for the DM evaporation is highly suppressed. Furthermore, compared to the sun, neutron stars have much higher density and deeper gravitational wells, so it is much more difficult to accelerate trapped DM above the escape velocity through interactions with neutrons. To evaporate from the neutron star, the DM has to gain enough energy such that its velocity is larger than the escape velocity of the neutron star. Because of the degeneracy effect, the typical energy transfer from the free neutron is $\sim T$; so the evaporation effect is relevant only when the DM mass is less than $\sim2T/v^2_{esc}\sim48~{\rm eV}$ for $T=10^5~{\rm K}$, which is much below the lower mass limit of our constraints $\sim2~{\rm keV}$ in the most optimistic case. Therefore, we can safely ignore the evaporation process.

\section{Asymmetric Scalar Dark Matter in Neutron Stars}
\subsection{Thermalization}
When DM particles are captured by the neutron star they lose energy via scattering with neutrons, and soon attain thermal equilibrium with the star. To estimate the thermalization time scale we calculate the DM energy loss rate:
\begin{equation}\label{dEdt}
\frac{dE}{dt}=- \xi n_B\sigma_n v~\delta E,
\end{equation}
where $n_B$ is the neutron number density in the center of the neutron star, $\delta E$ is the energy loss of the DM particle during each scattering event, and we use $\xi$ defined as in Eq.~\eqref{saturation} to parameterize the neutron degeneracy effect on the DM thermalization process. The typical velocity and the momentum transfer $\delta p= \sqrt{2} m_r v$ fully determine $\xi$. However, unlike the capture case, where the velocity is set by the escape velocity $v_{esc}$, the thermal equilibrium of DM particles and neutrons now sets $v\sim\sqrt{2E_{th}/\mx}$, where $E_{th}\simeq3T/2$ is the energy after thermalization. In the case of $\mx\gtrsim 1~{\rm GeV}$, $\delta p\sim\sqrt{2}m_Bv\simeq2.1\times10^{-5}~{\rm GeV}(T/10^5~{\rm K})^{1/2}(100~{\rm GeV}/\mx)^{1/2}$, which is much smaller than $p_F\simeq0.575~{\rm GeV}$. For $\mx \lesssim1~{\rm GeV}$, the momentum transfer is given by $\delta p\sim\sqrt{2}\mx v\simeq6.8\times10^{-5}(T/10^5~{\rm K})^{1/2}(\mx/0.1~{\rm GeV})^{1/2}$, which is again less than $p_F$. Therefore, the neutron degeneracy effect  reduces the DM thermalization efficiency over the entire DM mass range, and $\xi$ is everywhere given by $\xi\simeq\delta p/p_F$.

To estimate the thermalization time scale, we solve Eq.~(\ref{dEdt}) and get
\begin{equation}
t_{th}\simeq\frac{\mx^{2}m_Bp_F}{4\sqrt{2}n_B\sigma_n m^3_r}\frac{1}{E_{th}}.
\end{equation}
In the limit of $\mx\gtrsim 1~{\rm GeV}$, the thermalization time scale can be further simplified to $t_{th}\simeq\sqrt{2}\mx^{2}p_F/(12m^2_Bn_B\sigma_nT)$.
Taking typical values, we see
\begin{equation}
t_{th}\simeq0.054{~\rm years~}\left(\frac{\mx}{100~{\rm GeV}}\right)^{2}\left(\frac{2.1\times 10^{-45}~{\rm cm^2}}{\sigma_n}\right)\left(\frac{10^5~{\rm K}}{T}\right).
\label{coolingheavy}
\end{equation}
If DM mass is less than $1~{\rm GeV}$, the thermalization time scale is given by
\begin{eqnarray}
t_{th}\simeq7.7\times10^{-5}~{\rm years}\left(\frac{0.1~{\rm GeV}}{\mx}\right)\left(\frac{2.1\times 10^{-45}~{\rm cm^2}}{\sigma_n}\right)\left(\frac{10^5~{\rm K}}{T}\right).
\label{coolinglight}
\end{eqnarray}
To derive constraints on scalar ADM from black hole formation, we will assume the captured scalar ADM follows the thermal distribution in the neutron star. This is only true when $t_{th}$ is less than the neutron star age $\sim10^{10}~{\rm years}$. As we can see from Eqs.~(\ref{coolingheavy}) and (\ref{coolinglight}), light DM easily satisfies this condition. For heavy DM, $t_{th}$ is not always less than the neutron star age. In the following discussion we first assume the DM reaches thermal equilibrium with neutrons, and then we check the consistency of this assumption.

After attaining thermal equilibrium, captured DM particles drift to the center of the star and form an isothermal distribution with the typical radius
\begin{equation}
r_{th}=\left(\frac{9T}{4\pi G \rho_B \mx}\right)^{1/2}
\simeq24~{\rm cm}\left(\frac{T}{10^{5}~K} \cdot \frac{100~{\rm GeV}}{\mx}\right)^{1/2}.
\label{eq:rth}
\end{equation}
We can see that the captured DM particles very quickly occupy a very small region near the neutron star core.

\subsection{Self-gravitation and Black Hole Formation}

If the DM density is larger than the baryon density within the thermal radius $r_{th}$, the DM particles can become self-gravitating. For a total DM mass $M_X=N_X\mx$ within a thermal radius $r_{th}$, this condition is
\begin{equation}
\frac{3M_X}{4\pi r^3_{th}}\gtrsim\rho_B.
\end{equation}
Therefore, the DM becomes self-gravitating once the total number of DM particles is larger than a critical number
\begin{equation}
N_{self}\simeq4.8\times10^{41}\left(\frac{100~{\rm GeV}}{\mx}\right)^{5/2}\left(\frac{T}{10^5~{\rm K}}\right)^{3/2}.
\label{self}
\end{equation}
Recall the upper limit for the bosonic system given in Eq. (\ref{bosonlimit}) above which the zero point energy cannot prevent  gravitational collapse
\begin{equation}
N^{boson}_{Cha}\simeq1.5\times10^{34}\left(\frac{100~{\rm GeV}}{\mx}\right)^2.
\end{equation}
Thus, if the scalar ADM thermalizes and the mass satisfies $m_X\lesssim 10^{17}~{\rm GeV}\left(T/10^5~{\rm K}\right)^3$, we always have $N_{self}\gtrsim N^{boson}_{Cha}$. In this case, gravitational collapse occurs as soon as DM particles become self-gravitating in neutron stars.

\subsection{Bose-Einstein Condensation}
In the above discussion, we implicitly assumed that all captured scalar ADM particles followed a Maxwellian velocity distribution. At the extreme densities we are considering here, however, this minimal assumption is not necessarily satisfied. In particular, ensembles of bosonic particles at high densities exhibit novel statistical properties. If the central temperature of the neutron star falls below the critical temperature to form a Bose-Einstein condensate (BEC), the particles in the ground state condense and no longer follow the thermal distribution. We will now show that for light ADM this condensation increases the density and reduces the restriction on self-gravitation to such an extent that the number of ADM particles necessary for self-gravitation is less than the bosonic Chandrasekhar limit. Thus, gravitational collapse is set by $N^{boson}_{Cha}$.

To check this sequence of events, we begin by noting that for a given bosonic DM number density $n_X$, the critical temperature to form a BEC is given by
\begin{equation}
T_{c}=\frac{2\pi}{\mx}\left[\frac{n_X}{\zeta(3/2)}\right]^{2/3},
\end{equation}
where $\zeta$  is the Riemann-Zeta function, $\zeta(3/2)\simeq2.612$, and $n_X=3N_X/(4\pi r^3_{th})$. To see how likely it is that the captured ADM will form a BEC in the neutron star, we can estimate the critical ADM number as
\begin{eqnarray}
N_X=\zeta\left(\frac{3}{2}\right)\left(\frac{\mx T}{2\pi}\right)^{3/2}\frac{4\pi r^3_{th}}{3}\simeq1.0\times10^{36}\left(\frac{T}{10^5~{\rm K}}\right)^3,
\end{eqnarray}
where we have used Eq.~(\ref{eq:rth}). Therefore, if the total number of captured ADM in the neutron star is larger than $1.0\times10^{36}\left(T/10^5~{\rm K}\right)^3$, some of captured ADM particles will go to the ground state and form a BEC. This condition can be satisfied for a neutron star with relatively low central temperature as indicated by Eqs.~(\ref{NXlarge}) and (\ref{NXsmall}) .

For $T<T_c$, the BEC forms and the number of particles in the condensed ground state is
\begin{equation}
N^{0}_X=N_X\left[1-\left(\frac{T}{T_{c}}\right)^{3/2}\right]
\simeq N_X-1.0\times10^{36}\left(\frac{T}{10^5~{\rm K}}\right)^3.
\label{bec1}
\end{equation}
Since these ground-state particles effectively have zero temperature, they sink deep into the core of the neutron star. We can estimate the radius of  distribution of the ground state by requiring the zero point energy equal the gravitational energy
\begin{equation}\label{rbec}
 r_{BEC}=\left(\frac{3}{8\pi G \mx^2\rho_B}\right)^{1/4}
\simeq1.5\times10^{-5}~{\rm cm}\left(\frac{100~{\rm GeV}}{\mx}\right)^{1/2}.
\end{equation}
This is much smaller than $r_{th}$, which indicates a much higher DM density. Thus, the ground state itself may become self-gravitating. The critical number for the self-gravity of the DM particles in the condensed state is
\begin{equation}
N^{0}_{self}=\frac{4\pi}{3}\frac{\rho_Br^3_{BEC}}{\mx}
\simeq1.0\times10^{23}\left(\frac{100~{\rm GeV}}{\mx}\right)^{5/2}.
\end{equation}
Once the number of DM particles in the ground state is larger than $N^0_{self}$, these ground-state particles become self-gravitating. Since $N^0_{self}$ is less than $N_{self}$, the onset of self-gravity is marked by $N^0_{self}$ instead of $N_{self}$ in conditions where a BEC forms. As indicated above, this leads to qualitatively different behavior as compared to the case when a BEC does not form, since now $N^0_{self}$ can be less than $N^{boson}_{Cha}$.
If this is the case, as soon as a condensed ADM system reaches the Chandrasekhar limit it will undergo gravitational collapse.

For this effect to be important, $N^0_{X}$ has to grow larger than the Chandrasekhar limit for a bosonic system, so that the condition for black hole formation of the BEC becomes $N^0_X\gtrsim N^{boson}_{Cha}$. By using Eq.~(\ref{bec1}), we get a lower limit on the total DM number $N_X$,
 \begin{eqnarray}
\nonumber N_{BEC} &=& N^{boson}_{Cha}+1.0\times10^{36}\left(\frac{T}{10^{5}~{\rm K}}\right)^3\\
&\simeq&1.5\times10^{34}\left(\frac{100~{\rm GeV}}{\mx}\right)^2+1.0\times10^{36}\left(\frac{T}{10^{5}~{\rm K}}\right)^3.
\label{condbec}
\end{eqnarray}
We can see that the value of the right-hand side of Eq.~(\ref{condbec}) is less than $N_{self}$ if the DM mass $\mx\lesssim1.9\times10^{4}~{\rm GeV}$ ($4.7\times10^{3}~{\rm GeV}$) for a central temperature $T=10^5~{\rm K}$ ($10^6~{\rm K}$). In the situation where this condensation can occur, the BEC shortens the time scale for the black hole formation for the scalar ADM in the low-mass range.

\section{Hawking Radiation and Destruction of the Host Star}

During the collapse process, the gravitational contraction releases energy which can be absorbed by neutrons through DM-neutron scattering. This cooling mechanism is so efficient that eventually the DM sphere collapses to a black hole~\cite{Goldman:1989nd}. Once a black hole is formed at the center of the neutron star, it will rapidly capture the baryonic matter of the neutron star. Hawking radiation will also be active, reducing the mass of the black hole and possibly heating the remaining DM. Finally, the black hole may also consume the ambient DM particles, which can be crucial for the stability of the black hole. Here, we analyze the relative contributions of these effects, and we see that in the majority of our parameter space the black hole will grow and eventually consume the host neutron star. Once the physics of the accretion is properly considered, we find that Hawking radiation could be important for high- and intermediate-mass ADM.

For a black hole with mass $M_{BH}$, the differential equation that governs the rate of change of mass is
\begin{equation}
\frac{dM_{BH}}{dt}\simeq 4\pi \lambda_s\left(\frac{GM_{BH}}{v^2_s}\right)^2\rho_Bv_s-\frac{1}{15360\pi G^2 M^2_{BH}}+\left(\frac{dM_{BH}}{dt}\right)_{DM}.
\label{bondi}
\end{equation}
The second term of Eq.~(\ref{bondi}) represents the Hawking radiation rate, while the third term is the accretion rate of ambient DM particles. The first term of right-hand side of Eq.~(\ref{bondi}) is the Bondi-Hoyle accretion rate, in which $v_s=\sqrt{dP/d\rho}$ is the sound speed and $\lambda_s$ is the accretion eigenvalue for the transonic solution. To determine $v_s$ and $\lambda_s$, we characterize the equation of state of the neutrons by $P=K\rho^\gamma$, where $K$ and $\gamma$ are constant. For a nonrelativistic degenerate neutron gas, which is a good approximation for neutrons in the neutron star, we have $\gamma=5/3$ and $K=3^{2/3}\pi^{4/3}/(5m^{8/3}_B)$ \cite{Shapiro:1983du}. We estimate the sound speed as $v_s=\sqrt{K\gamma\rho^{\gamma-1}}\sim10^{5}~{\rm km/s}$, where we take $\rho\sim1.4\times10^{15}~{\rm g/cm^3}$. The accretion constant is given by $\lambda_s=(1/2)^{(\gamma+1)/(2\gamma-2)}[(5-3\gamma)/4]^{-(5-3\gamma)/(2\gamma-2)}=0.25$ \cite{Shapiro:1983du}.

\subsection{Black Hole Mass without Bose-Einstein Condensation}
For large ADM mass, we found above that $N_{self}<N_{BEC}$ and a black hole forms without the assistance of a BEC. After formation of the black hole, the neutron star continues capturing DM particles. These newly captured DM particles eventually sink to the center of the neutron star and distribute themselves within $r_{th}$. In principle, the black hole can increase its mass by capturing these additional DM particles. However, we find that this capture rate is very small and $(dM_{BH}/dt)_{DM}$ is negligible. This is because, for a nonrelativistic particle moving towards the black hole, its impact parameter must be less than $b_{max}=4GM_{BH}/v_\infty$~\cite{Shapiro:1983du} to penetrate the angular momentum barrier and fall into the black hole. Here, $v_\infty$ is the particle's velocity when it is far away from the black hole.

Taking $M_{BH}\sim \mx N_{self}$ and $v_\infty\sim\sqrt{3T/\mx}$, we can estimate $b_{max}$ as
\begin{equation}
b_{max}\sim1.6\times10^{-5}~{\rm cm}\left(\frac{10~{\rm TeV}}{\mx}\right)\left(\frac{T}{10^5}\right),
\end{equation}
which is much smaller than the thermal radius $r_{th}\simeq2.4~{\rm cm}(T/10^5~K)(10~{\rm TeV}/\mx)^{1/2}$. Therefore the majority of DM particles captured after formation of the black hole do not fall directly into the black hole. The remaining DM particles orbit the black hole at a distance of order $r_{th}$; the black hole gains mass from these particles at a rate set by the collisionless spherical accretion approximation~\cite{Shapiro:1983du}. We find that the DM accretion rate is much less than the baryon accretion rate, so we can safely ignore the $(dM_{BH}/dt)_{DM}$ term in this case, and we obtain a critical initial black hole mass
\begin{equation}
M^{crit}_{BH}\simeq1.2\times10^{37}~{\rm GeV}.
\end{equation}
Without a BEC, the initial black hole mass is $M_{BH}\sim N_{self}\mx$. If we demand $N_{self}\mx\gtrsim M^{crit}_{BH}$, we find that for $\mx \lesssim2.6\times10^6~{\rm GeV}\left(T/{10^5~{\rm K}}\right)$ the Hawking radiation has a longer time scale than the accretion process. Hence in this mass range the black hole will continuously accrete baryonic matter until the neutron star is consumed entirely.

\subsection{Black Hole Mass with Bose-Einstein Condensation}
For low-mass ADM, particles in the BEC ground state form a black hole. We must check the mass above which the black hole evaporates. If we naively ignore the term $(dM_{BH}/dt)_{DM}$ and demand $M_{BH}\sim \mx N^{boson}_{Cha}\gtrsim M^{crit}_{BH}$, we find that the black hole mass increases only for the DM mass less than $\sim 13~{\rm GeV}$. If so, the constraint is valid for $\mx\lesssim13~{\rm GeV}$. But in contrast to the non-BEC case, $(dM_{BH}/dt)_{DM}$ may have an important effect on the black hole mass evolution, and we find that the bound can be sensitive to masses higher than $\sim 13~{\rm GeV}$. We detail our reasoning below.

Since the black hole forms only from ADM particles in the ground state, the remaining ADM particles follow an isothermal distribution with a radius $r_{th}$. As discussed above, the thermally distributed DM particles do not fall into the black hole, and so the phase space of the non-BEC state is still completely occupied. Hence, if any more ADM particles are introduced to the thermal region, a new BEC ground state must form in the center of the star. In this way, the introduction of more DM particles into the thermal radius essentially forces the formation of a BEC ground state. Before and after the mini black hole forms, the neutron star continuously captures ADM particles. All of the captured ADM particles will eventually thermalize, sink to the center of the neutron star, and prompt the formation of a new BEC state.

If the thermalization time scale is shorter than the evaporation time scale of the mini black hole, the black hole can always efficiently accrete ADM particles in the new BEC state. Taking the initial black hole mass as $M_{BH}\sim \mx N^{boson}_{Cha}=1/(G\mx)$, we can conservatively estimate the black hole evaporation time scale in the absence of particle accretion as $t_{haw}\simeq15360\pi G^2M^3_{BH}/3\simeq5\times10^4~{\rm years}~(100~{\rm GeV}/\mx)^3$. This is much longer than the thermalization time scale given in Eq.~(\ref{coolingheavy}). Since all newly captured ADM particles eventually go to the ground state after the amount of time it takes them to thermalize, the rate at which ADM particles fall into the BEC ground state per unit time is given by the capture rate $C_B$. To check whether or not the ADM particles in the BEC state feed the black hole efficiently we calculate the maximal impact parameter $b_{max}$ and compare it with the distribution radius $r_{BEC}$.
With $M_{BH}\sim1/(G\mx)$ and $v_\infty\sim1/(\mx r_{BEC})$, we have
\begin{equation}
b_{max}\sim 4r_{BEC},
\end{equation}
where $r_{BEC}$ is given by Eq.~(\ref{rbec}). Since  $b_{max}\gtrsim r_{BEC}$, we see that the black hole can efficiently consume the BEC.  This occurs at a rate given by $(dM_{BH}/dt)_{DM}\sim \mx C_B$:
\begin{equation}
\left(\frac{dM_{BH}}{dt}\right)_{DM}\simeq2.3 \times 10^{36}~{\rm GeV/year}~ \left(\frac{\rho_X}{10^3~{\rm GeV/cm^3}}\right)\left(\frac{\sigma_{XB}}{2.1\times10^{-45}~{\rm cm^2}}\right).
\end{equation}
This new source of accretion overwhelms the Hawking radiation, which is emitted at the rate $\left(dM_{BH}/dt\right)_{hw}\sim10^{27}~{\rm GeV/year}~ (\mx/100~{\rm GeV})^2$. Thus, in the BEC case our constraints can be sensitive to scalar ADM with mass much higher than the naive estimate, $\mx \sim13~{\rm GeV}$.

In this discussion, we have assumed that the black hole formation and Hawking radiation do not destroy the BEC state. However, one may note that the black hole radiates as a blackbody at a very high temperature, $T_{haw} = (8\pi G M_{BH})^{-1} \sim \mx/8\pi$. If this Hawking radiation is emitted solely as relativistic particles of the standard model such as photons and neutrinos, these particles will heat the neutrons which in turn heat the ADM. We may estimate the change in neutron temperature by assuming that all of the initial rest mass energy of the black hole goes into heat. We see that in this case equipartition of energy requires that $\Delta T \sim  \mx N^{boson}_{Cha}/N_B \simeq 1.0\times 10^{-7} {\rm~K} ~({\rm 10~GeV}/\mx)$, and there is no change in the thermal ADM distribution.

On the other hand, if the energy produced by the Hawking radiation is efficiently transferred to DM particles, the situation may change. For example, the DM may couple to some light mediator particles, which can be produced by the Hawking radiation. These light mediators may heat the ADM directly and transfer the black hole energy to the thermal energy of ADM particles. In this case, the calculation is rather complicated and a detailed analysis is beyond the scope of the current paper. Here we give a conservative estimate by assuming that all of the initial black hole mass goes to the thermal energy of DM particles. The equipartition of energy gives $\Delta T \sim \mx N^{boson}_{Cha}/ N_{th} \simeq 1.7 \times 10^{14}~{\rm K}~(10~{\rm GeV}/\mx)\left(T/10^{5}~{\rm K}\right)^{-3}$, where the number of remaining non-BEC particles is $N_{th}=1.0\times10^{36}\left(T/10^{5}~{\rm K}\right)^3$, from Eq. \eqref{condbec}. This heating will greatly affect the thermal distribution of the scalar ADM particles. Since $r_{th} \propto \sqrt{T}$, the phase space will expand greatly, and the newly captured ADM particles do not form the BEC state. Thus, the black hole constraints are lifted in this case. 

In practice, for heating processes to occur, ADM particles must couple to light mediator particles, since the black hole can only produce particles with mass much less than $T_{haw}\sim\mx/8\pi$. On the other hand, the presence of the light mediator may also give rise to DM self-interactions, and the observed ellipticity of DM halos places a lower bound on the mediator mass~\cite{Feng:2009hw}. As an example, if $\mx\sim13~{\rm GeV}$, the halo shape constraint requires the mediator mass be larger than $\sim 40~{\rm MeV}$~\cite{Lin:2011gj}. Thus, for scalar ADM with mass $m_X\sim13~{\rm GeV}$ the existence of a mediator with mass in the range $\sim 40~{\rm MeV}-0.5~{\rm GeV}$ may help in evading the black hole formation constraints, while remaining consistent with the halo shape bound. We can see that the constraints for $\mx \gtrsim 13~{\rm GeV}$ in the BEC case are rather model-dependent, and we specify these regions in our plots.

Now we estimate the destruction time scale. If the initial black hole mass exceeds the critical value $M^{crit}_{BH}$, the time scale to destroy the neutron star is set by the Bondi-Hoyle rate
\begin{equation}
 t\sim\frac{v^3_s}{ \pi G^2\rho_B M_{BH}}
\simeq2.3\times10^{-5}~{\rm s}\left(\frac{M_{\odot}}{M^{i}_{BH}}\right),
\end{equation}
where we take $v_s\sim10^5~{\rm km/s}$ and $M_{\odot}$ is the mass of sun. Therefore, the characteristic time scale is $t\sim 17 ~{\rm years}\left(\mx/100~{\rm GeV}\right)^{3/2}(10^5~{\rm K}/T)^{3/2}$ and $5.4\times10^{6}~{\rm years}~(\mx/{\rm GeV})$ for the non-BEC case and BEC case, respectively, which we note is much shorter than the typical old neutron star age $\sim10^{10}~{\rm years}$.

\section{Observational Constraints}
So far, we have gone through the conditions for the captured scalar ADM particles to form a black hole. Such a black hole can destroy the host neutron star. We can see that these conditions are easily satisfied if the ADM particles have a sizable scattering cross section with neutrons. However, we observe many old neutron stars near the solar system and in globular clusters. Therefore, these observed old neutrons stars constrain the ADM-neutron scattering cross section. In this section, we proceed to derive upper bounds on the ADM-neutron scattering cross section for a given ADM mass. In Secs.~\ref{sec:VIA} (\ref{sec:VIB}) we first derive constraints on $\sigma_n$ by assuming fiducial neutron star parameters for the case without (with) Bose-Einstein condensation. In Sec.~\ref{sec:VIC}, we apply these bounds to observed neutron stars in the solar neighborhood and in the globular cluster M4.

\subsection{Constraints on DM-neutron cross section without Bose-Einstein condensation}\label{sec:VIA}
We first discuss constraints in the absence of BEC formation; this gives rise to conservative constraints in the low-mass range but is more stringent for high masses, where BECs do not readily form. Since in this case $N_{self}>N^{boson}_{Cha}$ for $\mx\lesssim10^{17}~{\rm GeV}(T/10^{5}~{\rm K})^3$, gravitational collapse occurs as soon as DM particles start self-gravitating. In order to avoid the destruction of neutron stars, we demand $N_X<N_{self}$ and get an upper bound on DM-neutron scattering cross section $\sigma_{n}$. For DM with mass $m_X \gtrsim 1~{\rm GeV}$ we derive the following bound on $\sigma_n$ using Eqs.~(\ref{NXlarge}) and (\ref{self}) and requiring that the host neutron star is not destroyed over the course of its lifetime $t$
\begin{equation}
\sigma_{n}<4.4\times10^{-48}~{\rm cm^2}
\left(\frac{10^{3}~{\rm GeV/cm^3}}{\rho_X} \cdot \frac{10^{10}~{\rm years}}{t}\right)
\left(\frac{100~{\rm GeV}}{\mx} \cdot \frac{T}{10^5~{\rm K}}\right)^{3/2}.
\label{boundheavy}
\end{equation}
For $\mx\lesssim 1~{\rm GeV}$, the bound on $\sigma_n$ is
\begin{equation}
\sigma_{n}<9.3\times10^{-46}~{\rm cm^2}
\left(\frac{10^{6}~{\rm GeV/cm^3}}{\rho_X} \cdot \frac{10^{10}~{\rm years}}{t}\right)
\left(\frac{0.1~{\rm GeV}}{\mx}\right)^{5/2}\left( \frac{T}{10^5~{\rm K}}\right)^{3/2}.
\label{boundlight}
\end{equation}
Since the capture rate saturates when the DM-neutron scattering cross section is larger than $\sigma_{max}\simeq2.1\times10^{-45}~{\rm cm^2}$, the upper bound on $\sigma_n$ is only valid when the value of the right-hand side of Eq.~(\ref{boundheavy}) and Eq.~(\ref{boundlight}) is smaller than $\sigma_{max}$. Because the capture and thermalization processes do not distinguish between spin-dependent and spin-independent cross sections, the bound on $\sigma_n$ applies to both cases.

\begin{figure}
    \includegraphics[width=0.49\textwidth]{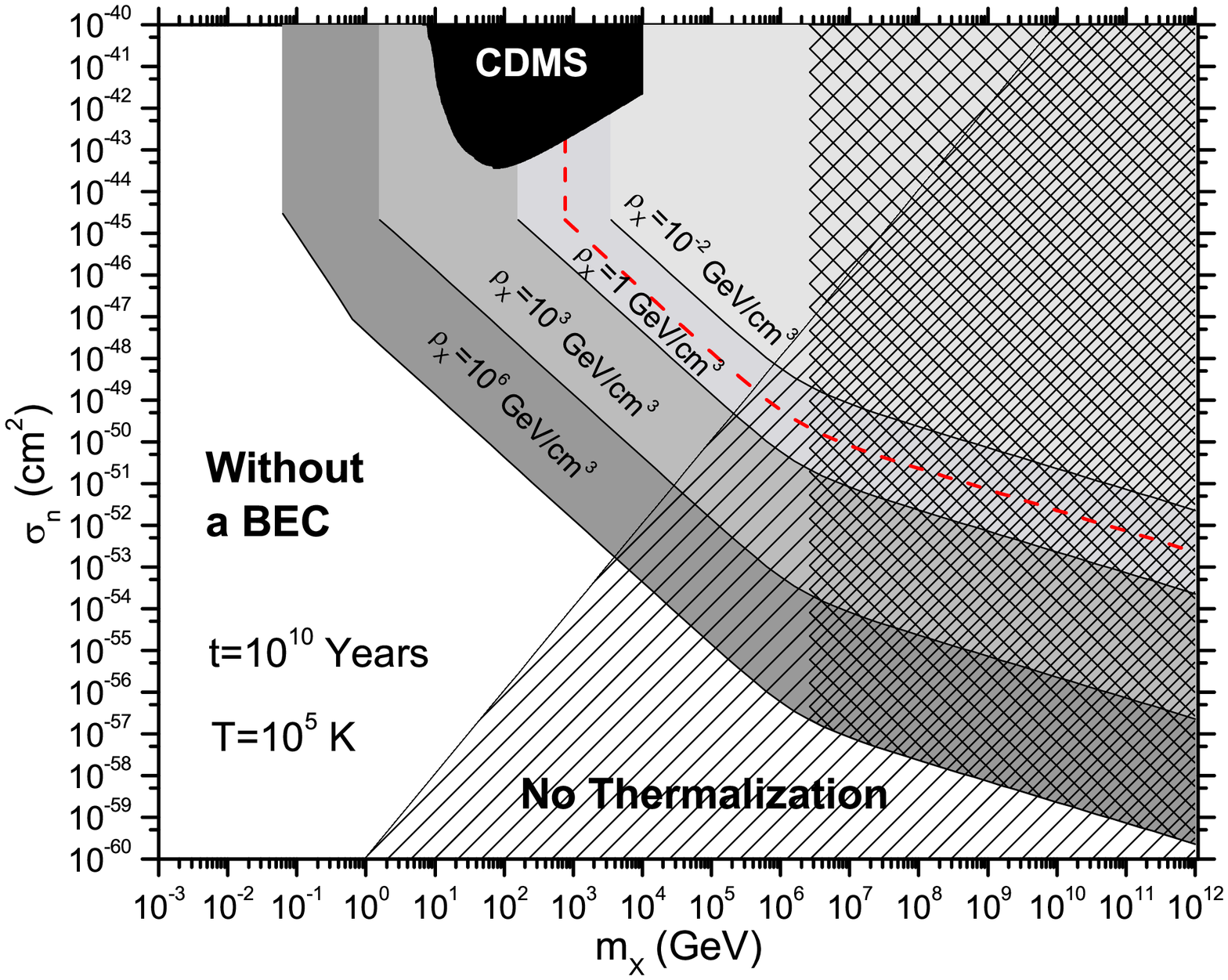}
      \includegraphics[width=0.49\textwidth]{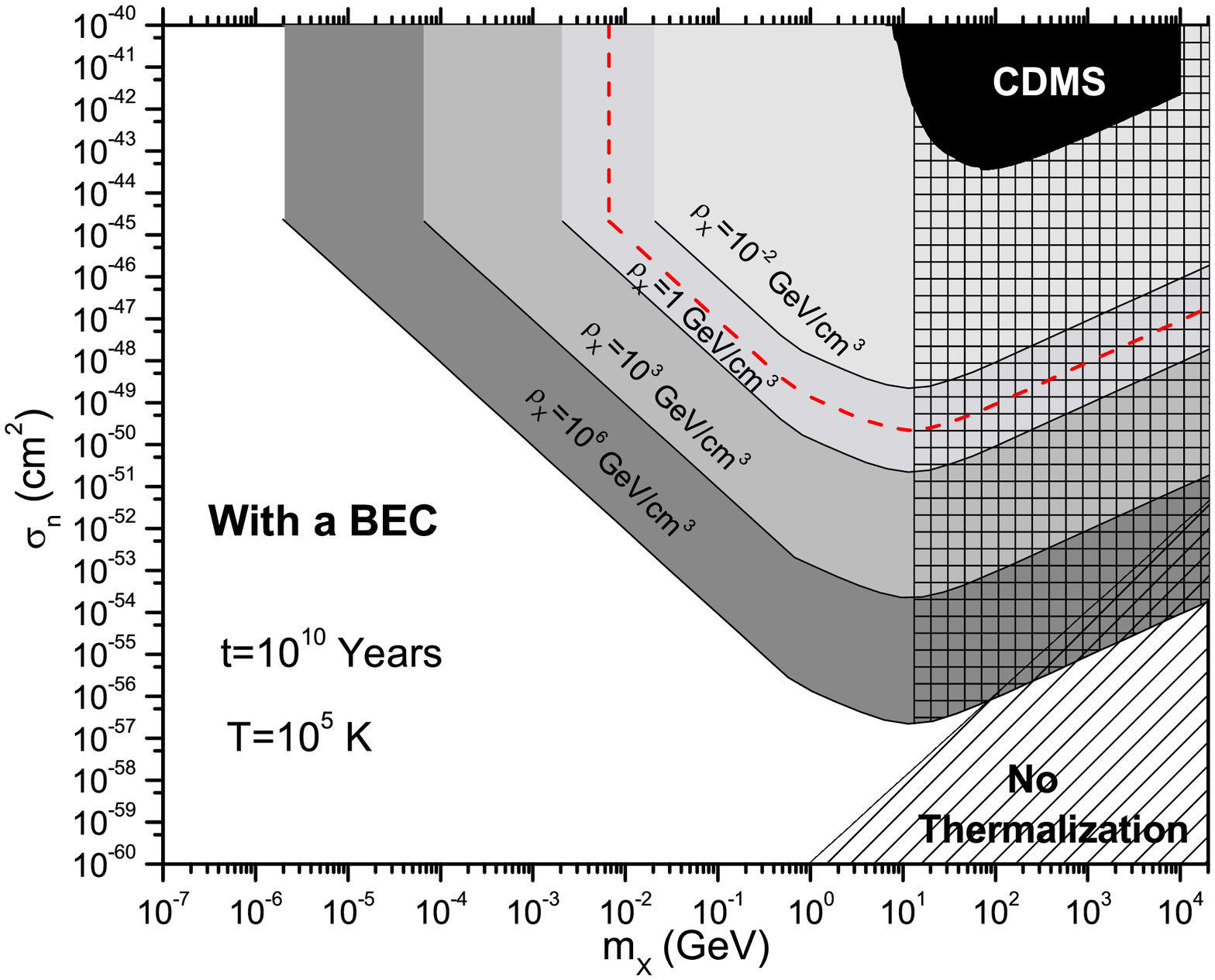}
     \caption{Regions (gray) of DM-neutron scattering cross section in which accumulated scalar ADM forms a black hole. We have $\rho_X=0.1~{\rm GeV/cm^3}$ along the dotted red lines, which is approximately the DM density in the solar neighborhood. In the left panel, we assume a BEC does not form and all captured DM particles become self-gravitating and collapse. In the right panel, we assume a BEC does form and DM particles in the BEC ground-state collapse and form a black hole. We take the neutron star age to be $10^{10}~{\rm years}$ with a central temperature $10^{5}~{\rm K}$. In the diagonally shaded regions, DM particles cannot thermalize with neutrons within the age of neutron star. In the diagonally cross-hatched region, the black hole can evaporate due to the Hawking radiation. In the square-hatched regions, Hawking radiation may interfere with DM accretion by heating the thermalized DM, and in some cases the black hole can evaporate (see discussion in Sec. V). In these hatched regions, the bounds are lifted. The black regions are excluded by recent CDMS results (spin-independent)~\cite{Ahmed:2009zw}.}
\label{bhexcl}
\end{figure}

We depict the constraints from requiring neutron star survival in the left panel of Fig.~\eqref{bhexcl} with various values of the DM density. We take the central temperature as $10^5~{\rm K}$ and neutron star age to be $10^{10}~{\rm years}$. The most prominent qualitative feature, the sharp vertical cutoff, corresponds to cross sections $\sigma_n\simeq2.1\times10^{-45}~{\rm cm^2}$. Here, the geometric cross section limits the capture of DM particles, so we cannot constrain the interaction cross section for the mass below the cutoff. Furthermore, without a BEC, we can constrain scalar ADM with mass $\mx \lesssim 100~{\rm MeV}$ only if the DM density $\rho_X \gtrsim 10^6~{\rm GeV/cm^3}$. One may find regions with such high DM density near the galactic center. As we discussed before, if the DM mass is less than $\sim 1~{\rm GeV}$, the capture rate is reduced due to the neutron degeneracy effect, which is indicated by the change in slope of the curve with $\rho_X=10^{6}~{\rm GeV/cm^3}$ when $\mx\lesssim 1~{\rm GeV}$. For DM mass $m_X\gtrsim10^6~{\rm GeV}$, the factor $B^2$ in Eq.~\eqref{rate} falls below 1 and the capture probability through a single scatter becomes lower, as shown by the bump in the left panel of the figure. In the hatched region, DM particles cannot thermalize with the surrounding neutrons within the age of neutron star as discussed in Sec. IV A. In this region, the captured DM does not necessarily distribute within the small thermal radius of the neutron star core, and the bound does not apply. In the cross-hatched region, the initial black hole mass is so small that it can evaporate due to the Hawking radiation as shown in Sec. V A. In the square-hatched region, the Hawking radiation may interfere with DM
accretion, and the black hole will evaporate in some case as discussed in Sec. V B.

We can also estimate the analogous bound from white dwarf stars. Typical white dwarf parameters are~\cite{Kepler:2006ns}: mass $\sim0.7~M_\odot$, radius $\sim6.3\times10^3~{\rm km}$, density $\sim10^{6}~{\rm g/cm^3}$, central temperature $\sim10^7~{\rm K}$ and escape velocity $\sim 6\times10^3~{\rm km/s}$. White dwarfs are composed of carbon and oxygen; we make the conservative assumption that the white dwarf is entirely composed of carbon. We find that the white dwarf bound on $\sigma_n$ is about 9 orders of magnitude weaker than the limit derived for typical neutron stars.

\subsection{Constraints on DM-neutron cross section with Bose-Einstein condensation}\label{sec:VIB}

If the captured scalar ADM does form a BEC, the constraints on the DM-neutron cross section become stronger. This is because $N_{BEC} \lesssim N_{self}$ for the DM mass less than a few TeV, dependent on the central temperature $T$. In this case, self-gravitation and gravitational collapse will occur more quickly due to the heightened density of the ADM.

For $\mx\gtrsim1~{\rm GeV}$, the requirement $N_X\lesssim N_{BEC}$ gives
\begin{align}
\nonumber\sigma_n & \lesssim  9.1\times10^{-54}~{\rm cm^2}\left(\frac{10^{3}~{\rm GeV/cm^3}}{\rho_X} \cdot \frac{10^{10}~{\rm years}}{t}\right)\left(\frac{\mx}{100~{\rm GeV}}\right)\\& \qquad \times\left[1.5\times10^{-2}\left(\frac{100~{\rm GeV}}{\mx}\right)^2+\left(\frac{T}{10^{5}~{\rm K}}\right)^3\right].
\end{align}
In the case of $\mx\lesssim1~{\rm GeV}$, the upper limit on $\sigma_n$ is given by
\begin{align}
\nonumber\sigma_n & \lesssim 6.2\times10^{-56}~{\rm cm^2}\left(\frac{10^{3}~{\rm GeV/cm^3}}{\rho_X} \cdot \frac{10^{10}~{\rm years}}{t}\right)\\& \qquad \times\left[1.5\times10^4\left(\frac{0.1~{\rm GeV}}{\mx}\right)^2+\left(\frac{T}{10^{5}~{\rm K}}\right)^3\right].
\end{align}
Again, we derive bounds by requiring that the host star is not destroyed by the gravitational collapse of the ADM.

In the right panel of Fig.~\eqref{bhexcl}, we display the DM-neutron scattering cross section with various values of the DM density that satisfy $N_X\gtrsim N_{BEC}$. For $m_X \lesssim 10 \mbox{ GeV}$, the BEC forms before the Chandrasekhar mass is reached, while for $m_X \gtrsim 10 \mbox{ GeV}$, the Chandrasekhar mass is reached before the BEC forms, so that collapse of the DM to a black hole occurs as soon as the BEC forms.  The change in the slope of the curves around $m_X \sim 1 \mbox{ GeV}$ is a combination of this effect with a decreased capture efficiency below $m_X \sim 1 \mbox{ GeV}$.  We can see the formation of the BEC significantly improves the bound for light ADM. As an example, we see that for $\rho_X=1{\rm ~GeV/cm^3}$, BEC formation strengthens the constraint as long as $\mx \lesssim 13~{\rm GeV}$, while for higher masses the entire mass of ADM becomes self-gravitating before BEC formation occurs.

\subsection{Constraints from observed pulsars}\label{sec:VIC}
Now we consider the observations of a few relatively cold and old neutron stars that can provide tests of this effect. PSR J0437-4715 is a nearby pulsar at a distance of about $139\pm3~{\rm pc}$ from the solar system. The surface temperature is $T_e=1.2\times10^5~{\rm K}$~\cite{Kargaltsev:2003eb}. When its secular motion is accounted for, calculations indicate its age is $6.69\times10^9~{\rm years}$~\cite{Manchester:2004bp}.  Another nearby pulsar is PSR J2124-3358, located $270~{\rm pc}$ away from us with a surface temperature $T_e<4.6\times10^5~{\rm K}$~\cite{Kargaltsev:2003eb}. Its age is $7.81\times10^9~{\rm years}$~\cite{Manchester:2004bp}.

For these nearby pulsars, we can calculate the central temperature from the surface temperature $T_e$ by using the analytical formula~\cite{Gundmundsson}
\begin{equation}
T\simeq1.288\times10^8~{\rm K}\left[\frac{10^{14}~{\rm cm/s}}{g_s}\left(\frac{T_e}{10^6~{\rm K}}\right)^4\right]^{0.455},
\end{equation}
where $g_s=GM_n/R^2_n$ is the surface gravity and we take $g_s\simeq1.7\times10^{14}~{\rm cm/s^2}$ for $M_n=1.44~M_\odot$ and $R_n=10.6~{\rm km}$. We find the central temperature is $2.1\times10^{6}~{\rm K}$ and $2.5\times10^7~{\rm K}$ for J0437-4715 and J2124-3358, respectively. We take the ambient DM density to be $0.3~{\rm GeV/cm^3}$, because these pulsars are in our relative neighborhood.
\begin{figure}
    \includegraphics[width=0.49\textwidth]{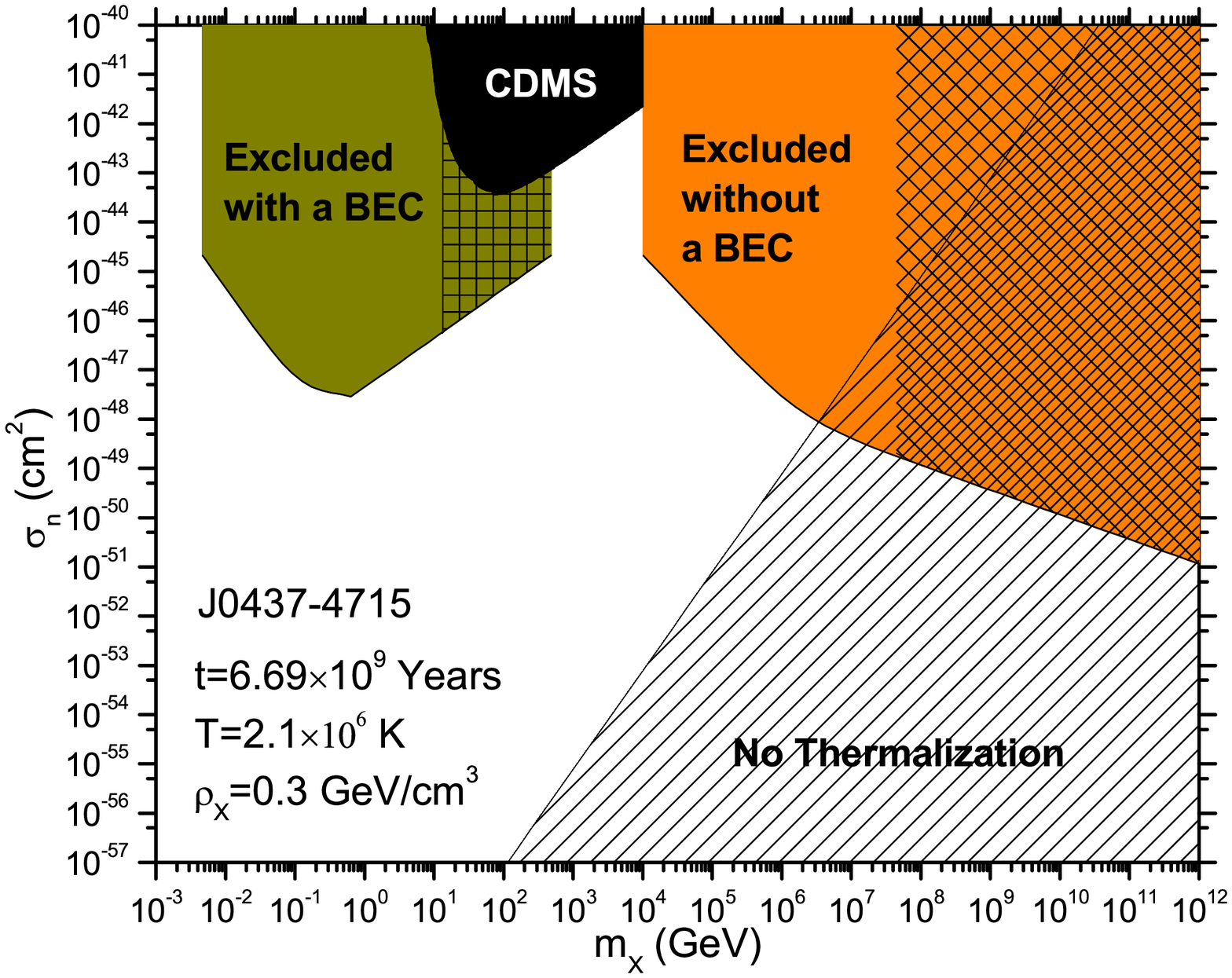}
    \includegraphics[width=0.49\textwidth]{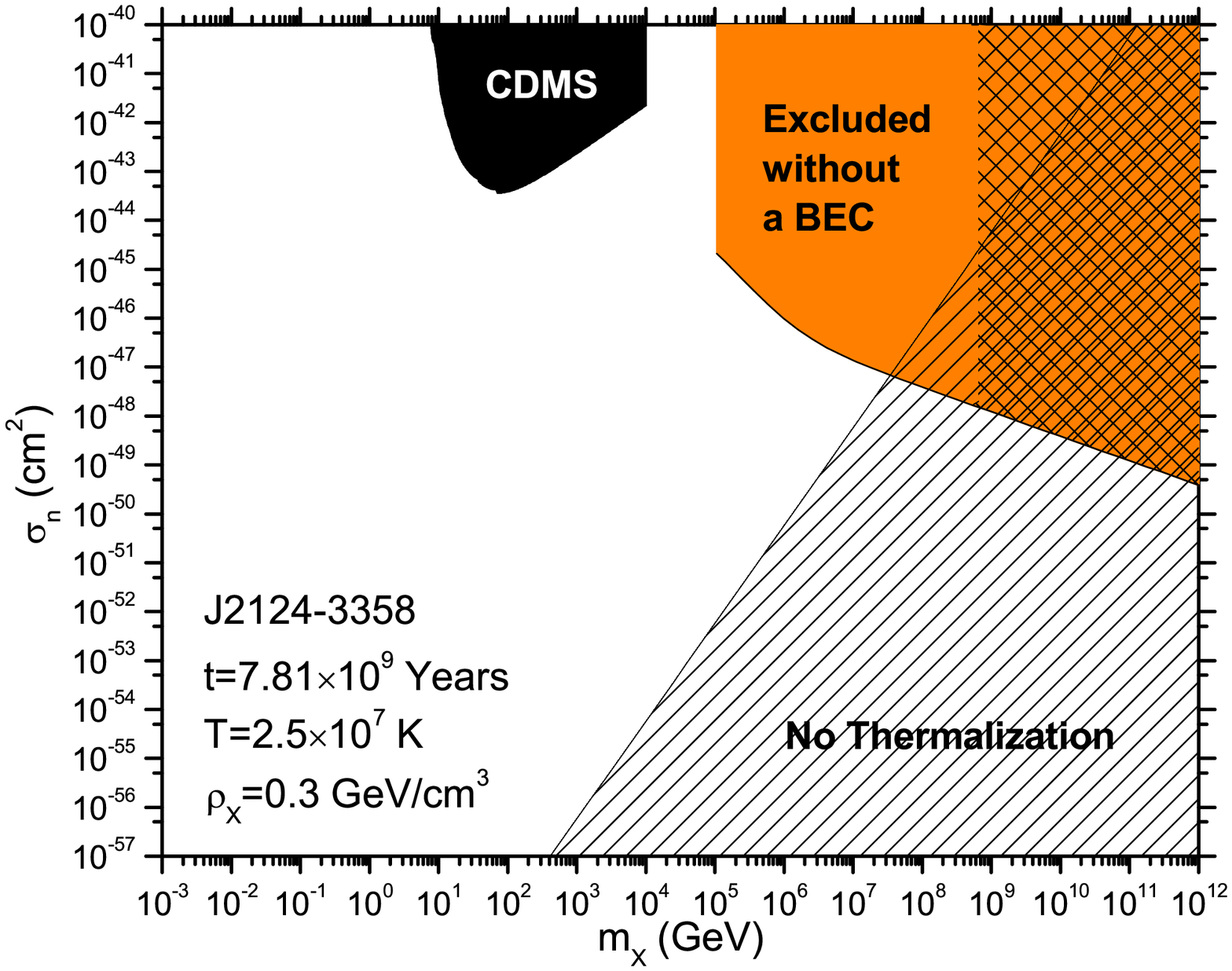}
    \caption{Regions (colored) excluded by the nearby pulsars J0437-4715 (left) and J2124-3358 (right). The shaded, diagonal and square cross-hatched, and black regions are as in Fig.~({\ref{bhexcl}}). In hatched regions, the bounds are lifted.}
\label{npls}
\end{figure}

In Fig.~(\ref{npls}), we show the constraints on the DM-nucleon scattering cross section of the scalar ADM from the nearby pulsars J0437-4715 (left panel) and J2124-3358 (right panel). We can see that J0437-4715 can constrain scalar ADM with $m_X \gtrsim 10~{\rm TeV}$ without a BEC. With the formation of a BEC, it is also sensitive to the mass range $ m_X \sim 5~{\rm MeV}-13~{\rm GeV}$. The captured scalar ADM cannot form a BEC in the pulsar J2124-3358. This is because it has a relatively high central temperature, and the formation of a BEC requires a DM-nucleon cross section larger than the saturation cross section $\sigma_{max}\simeq2.1\times10^{-45}~{\rm cm^2}$.

\begin{figure}
    \includegraphics[width=0.49\textwidth]{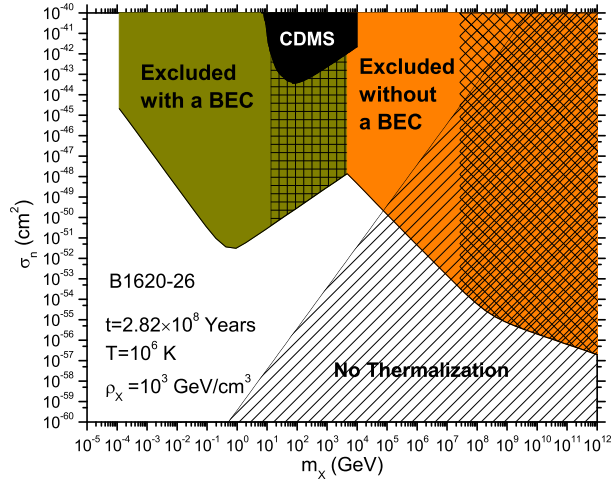}
    \caption{Regions (colored) excluded by the pulsar B1620-26 in the globular cluster M4. Note the globular cluster M4 is a baryon-dominated system, and there may in fact be no dark matter in this system. Here, we take $\rho_X=10^{3}~{\rm GeV/cm^3}$ motivated by numerical results in Refs.~\cite{Bertone:2007ae,McCullough:2010ai}. The shaded, diagonal and square cross hatched, and black regions are as in Fig.~({\ref{bhexcl}}). In hatched regions, the bounds are lifted.}
\label{m4}
\end{figure}

Since the bound is sensitive to the DM density, we also consider neutron stars in regions with high $\rho_X$.  Globular clusters possibly offer this type of environment, and observations of Pulsar B1620-26 place it in the globular cluster M4~\cite{globular} with an age of $2.82\times10^{8}~{\rm years}$ \cite{Kargaltsev:2003eb}. Since it is far away from us, its surface temperature is unknown, and we are not able to calculate its central temperature. In our analysis, we take $T=10^6~{\rm K}$ as a reasonable approximation due to its advanced age. We take $\rho_X=10^{3}~{\rm GeV/cm^3}$ for the DM density and $\bar{v}=20~{\rm km/s}$, motivated by discussions in Refs.~\cite{Bertone:2007ae,McCullough:2010ai}. Note that the exact value of DM density in globular clusters is unknown. Globular clusters are baryon-dominated systems, and currently there is no evidence that DM is present in these systems; see Ref.~\cite{Mashchenko:2004hj} for simulations of DM content in globular clusters.  In Fig.~(\ref{m4}), we show the constraints on the DM-nucleon scattering cross section of scalar ADM from the pulsar B1620-26 in the globular cluster M4. Note that when the DM mass is larger than $\sim 4.7\times10^3~{\rm GeV}$, $N_{BEC}\gtrsim N_{self}$ and all captured DM particles collapse before a BEC forms.

\section{Conclusions}
We have studied the consequences of scalar ADM accumulation in neutron stars. Neutron stars have high density and are ideal objects for capturing DM at high rates.  Since ADM does not self-annihilate, a high mass of DM can accrete in the neutron star, and, lacking Fermi degeneracy pressure, rapidly self-gravitate and exceed the Chandrasekhar limit. Furthermore, the formation of a BEC increases the density of the ADM by several orders of magnitude, which greatly accelerates the onset of gravitational collapse and considerably strengthens the constraints in certain regions of parameter space. For high dark matter mass, which corresponds to a low initial black hole mass, Hawking radiation can weaken the constraints. In the absence of light messenger particles, this effect is quite limited in scope, but if the black hole can heat the ADM directly it may affect a larger range of parameters.

We have computed the size of all of these effects and found that some presently observed pulsars constrain scalar ADM far more tightly than what is currently possible with direct detection experiments. These constraints are stronger even than the upcoming generation of experiments.  We also note that these constraints can be significantly improved in the future with observations of old pulsars in regions of DM density greater than $10^3 \mbox{ GeV/cm}^3$.
\\ \\
\noindent {\bf Note added:} Since the first version of our paper appeared, \cite{Kouvaris:2011fi} was written. In the case where a BEC forms and Hawking radiation is important, the authors of \cite{Kouvaris:2011fi} find that scalar ADM with mass $\sim0.75~{\rm MeV}-16~{\rm GeV}$ is excluded by nearby neutron stars if the scattering cross section with neutrons is sizable. Our results exclude scalar ADM in the mass range $\sim5~{\rm MeV}-13~{\rm GeV}$ in the same scenario. The discrepancy in the low-mass limit comes about because we take into account the neutron degeneracy effects on the capture rate. This is important for DM mass below $\sim1$ GeV.


\begin{thebibliography}{99}

\bibitem{review}
  G.~Jungman, M.~Kamionkowski, K.~Griest,
  Phys.\ Rept.\  {\bf 267}, 195-373 (1996).
  [hep-ph/9506380].
  G.~Bertone, D.~Hooper, J.~Silk,
  Phys.\ Rept.\  {\bf 405}, 279-390 (2005).
  [hep-ph/0404175].
  J.~L.~Feng,
  [arXiv:1003.0904 [astro-ph.CO]].


\bibitem{Early}
  S.~Nussinov,
  Phys.\ Lett.\  {\bf B165}, 55 (1985).
  S.~M.~Barr, R.~S.~Chivukula and E.~Farhi,
  Phys.\ Lett.\  B {\bf 241}, 387 (1990).
  S.~M.~Barr,
  Phys.\ Rev.\  D {\bf 44}, 3062 (1991).
  D.~B.~Kaplan,
  Phys.\ Rev.\ Lett.\  {\bf 68}, 741 (1992).

 \bibitem{outofequilibrium}
  R.~Kitano and I.~Low,
  Phys.\ Rev.\  D {\bf 71}, 023510 (2005)
  [arXiv:hep-ph/0411133].


\bibitem{Asymmetric}
   D.~E.~Kaplan, M.~A.~Luty and K.~M.~Zurek,
  Phys.\ Rev.\  D {\bf 79}, 115016 (2009)
  [arXiv:0901.4117 [hep-ph]].

\bibitem{models}
  H.~An, S.~L.~Chen, R.~N.~Mohapatra and Y.~Zhang,
  JHEP {\bf 1003}, 124 (2010)
  [arXiv:0911.4463 [hep-ph]].
  P.~H.~Gu,
  Phys.\ Rev.\  D {\bf 81}, 095002 (2010)
  [arXiv:1001.1341 [hep-ph]].
  B.~Feldstein and A.~L.~Fitzpatrick,
  JCAP {\bf 1009}, 005 (2010)
  [arXiv:1003.5662 [hep-ph]].
  J.~Shelton and K.~M.~Zurek,
  Phys.\ Rev.\  D {\bf 82}, 123512 (2010)
  [arXiv:1008.1997 [hep-ph]].
  H.~Davoudiasl, D.~E.~Morrissey, K.~Sigurdson and S.~Tulin,
  Phys.\ Rev.\ Lett.\  {\bf 105}, 211304 (2010)
  [arXiv:1008.2399 [hep-ph]].
  N.~Haba and S.~Matsumoto,
  arXiv:1008.2487 [hep-ph].
  E.~J.~Chun,
  arXiv:1009.0983 [hep-ph].
  P.~H.~Gu, M.~Lindner, U.~Sarkar and X.~Zhang,
  arXiv:1009.2690 [hep-ph].
  M.~Blennow, B.~Dasgupta, E.~Fernandez-Martinez and N.~Rius,
  arXiv:1009.3159 [hep-ph].
  J.~McDonald,
  arXiv:1009.3227 [hep-ph].
  S.~R.~Behbahani, M.~Jankowiak, T.~Rube and J.~G.~Wacker,
  arXiv:1009.3523 [hep-ph].
  L.~J.~Hall, J.~March-Russell and S.~M.~West,
  arXiv:1010.0245 [hep-ph].
  R.~Allahverdi, B.~Dutta and K.~Sinha,
  arXiv:1011.1286 [hep-ph].
  B.~Dutta and J.~Kumar,
  arXiv:1012.1341 [hep-ph].
  J.~J.~Heckman and S.~J.~Rey,
  arXiv:1102.5346 [hep-th].
  M.~T.~Frandsen, S.~Sarkar and K.~Schmidt-Hoberg,
  arXiv:1103.4350 [hep-ph].



\bibitem{Cohen:2010kn}
  T.~Cohen, D.~J.~Phalen, A.~Pierce and K.~M.~Zurek,
  Phys.\ Rev.\  D {\bf 82}, 056001 (2010)
  [arXiv:1005.1655 [hep-ph]].

\bibitem{Kang:2011wb}
  Z.~Kang, J.~Li, T.~Li, T.~Liu, J.~Yang,
  [arXiv:1102.5644 [hep-ph]].

\bibitem{Bernabei:2008yi}
  R.~Bernabei {\it et al.}  [DAMA Collaboration],
  Eur.\ Phys.\ J.\  C {\bf 56}, 333 (2008)
  [arXiv:0804.2741 [astro-ph]].

\bibitem{Aalseth:2010vx}
  C.~E.~Aalseth {\it et al.}  [CoGeNT collaboration],
  arXiv:1002.4703 [astro-ph.CO].

 \bibitem{quirky}
    D.~Hooper, J.~March-Russell, S.~M.~West,
  Phys.\ Lett.\  {\bf B605}, 228-236 (2005).
  [hep-ph/0410114].
  S.~B.~Gudnason, C.~Kouvaris, F.~Sannino,
  Phys.\ Rev.\  {\bf D74}, 095008 (2006).
  [hep-ph/0608055].
    R.~Foadi, M.~T.~Frandsen, F.~Sannino,
  Phys.\ Rev.\  {\bf D80}, 037702 (2009).
  [arXiv:0812.3406 [hep-ph]].
    G.~D.~Kribs, T.~S.~Roy, J.~Terning and K.~M.~Zurek,
  Phys.\ Rev.\  D {\bf 81}, 095001 (2010)
  [arXiv:0909.2034 [hep-ph]].
  M.~R.~Buckley and L.~Randall,
  arXiv:1009.0270 [hep-ph].
  N.~Haba, S.~Matsumoto and R.~Sato,
  arXiv:1101.5679 [hep-ph].
  E.~J.~Chun,
  arXiv:1102.3455 [hep-ph].
  M.~L.~Graesser, I.~M.~Shoemaker and L.~Vecchi,
  arXiv:1103.2771 [hep-ph].




\bibitem{Lepto}
  T.~Cohen and K.~M.~Zurek,
  Phys.\ Rev.\ Lett.\  {\bf 104}, 101301 (2010)
  [arXiv:0909.2035 [hep-ph]].

\bibitem{Falkowski:2011xh}
  A.~Falkowski, J.~T.~Ruderman, T.~Volansky,
  arXiv:1101.4936 [hep-ph].

\bibitem{Ahmed:2009zw}
  Z.~Ahmed {\it et al.}  [The CDMS-II Collaboration],
  Science {\bf 327}, 1619 (2010)
  [arXiv:0912.3592 [astro-ph.CO]].

\bibitem{Angle:2007uj}
  J.~Angle {\it et al.}  [XENON Collaboration],
  Phys.\ Rev.\ Lett.\  {\bf 100}, 021303 (2008)
  [arXiv:0706.0039 [astro-ph]].

\bibitem{Lebedenko:2008gb}
  V.~N.~Lebedenko {\it et al.},
  Phys.\ Rev.\  D {\bf 80}, 052010 (2009)
  [arXiv:0812.1150 [astro-ph]].

\bibitem{Beltran:2008xg}
  M.~Beltran, D.~Hooper, E.~W.~Kolb and Z.~C.~Krusberg,
  Phys.\ Rev.\  D {\bf 80}, 043509 (2009)
  [arXiv:0808.3384 [hep-ph]].

\bibitem{Cao:2009uw}
  Q.~H.~Cao, C.~R.~Chen, C.~S.~Li and H.~Zhang,
  arXiv:0912.4511 [hep-ph].

\bibitem{Goodman:2010yf}
  J.~Goodman, M.~Ibe, A.~Rajaraman, W.~Shepherd, T.~M.~P.~Tait and H.~B.~Yu,
  Phys.\ Lett.\  B {\bf 695}, 185 (2011)
  [arXiv:1005.1286 [hep-ph]].
  J.~Goodman, M.~Ibe, A.~Rajaraman, W.~Shepherd, T.~M.~P.~Tait and H.~B.~Yu,
  Phys.\ Rev.\  D {\bf 82}, 116010 (2010)
  [arXiv:1008.1783 [hep-ph]].

\bibitem{Bai:2010hh}
  Y.~Bai, P.~J.~Fox and R.~Harnik,
  JHEP {\bf 1012}, 048 (2010)
  [arXiv:1005.3797 [hep-ph]].
  P.~J.~Fox, R.~Harnik, J.~Kopp, Y.~Tsai,
[arXiv:1103.0240 [hep-ph]].


\bibitem{Press:1985ug}
  W.~H.~Press and D.~N.~Spergel,
  Astrophys.\ J.\  {\bf 296}, 679 (1985).

\bibitem{Gould:1987ju}
  A.~Gould,
  Astrophys.\ J.\  {\bf 321}, 560 (1987).
  A.~Gould,
  Astrophys.\ J.\  {\bf 321}, 571 (1987).

\bibitem{Griest:1986yu}
  K.~Griest and D.~Seckel,
  Nucl.\ Phys.\  B {\bf 283}, 681 (1987)
  [Erratum-ibid.\  B {\bf 296}, 1034 (1988)].

\bibitem{Spolyar:2007qv}
  D.~Spolyar, K.~Freese and P.~Gondolo,
  Phys.\ Rev.\ Lett.\  {\bf 100}, 051101 (2008)
  [arXiv:0705.0521 [astro-ph]].

\bibitem{Iocco:2008rb}
  F.~Iocco, A.~Bressan, E.~Ripamonti, R.~Schneider, A.~Ferrara and P.~Marigo,
 Mon.\ Not.\ Roy.\ Astron.\ Soc.\  {\bf 390}, 1655 (2008)
 [arXiv:0805.4016 [astro-ph]].

\bibitem{McCullough:2010ai}
  M.~McCullough and M.~Fairbairn,
  Phys.\ Rev.\  D {\bf 81}, 083520 (2010)
  [arXiv:1001.2737 [hep-ph]].

\bibitem{Hooper:2010es}
  D.~Hooper, D.~Spolyar, A.~Vallinotto and N.~Y.~Gnedin,
  Phys.\ Rev.\  D {\bf 81}, 103531 (2010)
  [arXiv:1002.0005 [hep-ph]].

\bibitem{Kouvaris:2007ay}
  C.~Kouvaris,
  Phys.\ Rev.\  D {\bf 77}, 023006 (2008)
  [arXiv:0708.2362 [astro-ph]].

\bibitem{Kouvaris:2010vv}
  C.~Kouvaris and P.~Tinyakov,
  Phys.\ Rev.\  D {\bf 82}, 063531 (2010)
  [arXiv:1004.0586 [astro-ph.GA]].

\bibitem{deLavallaz:2010wp}
  A.~de Lavallaz and M.~Fairbairn,
  Phys.\ Rev.\  D {\bf 81}, 123521 (2010)
  [arXiv:1004.0629 [astro-ph.GA]].

\bibitem{Frandsen:2010yj}
  M.~T.~Frandsen and S.~Sarkar,
  Phys.\ Rev.\ Lett.\  {\bf 105}, 011301 (2010)
  [arXiv:1003.4505 [hep-ph]].

\bibitem{Cumberbatch:2010hh}
  D.~T.~Cumberbatch, J.~A.~Guzik, J.~Silk, L.~S.~Watson and S.~M.~West,
  Phys.\ Rev.\  D {\bf 82}, 103503 (2010)
  [arXiv:1005.5102 [astro-ph.SR]].

\bibitem{Taoso:2010tg}
  M.~Taoso, F.~Iocco, G.~Meynet, G.~Bertone and P.~Eggenberger,
  Phys.\ Rev.\  D {\bf 82}, 083509 (2010)
  [arXiv:1005.5711 [astro-ph.CO]].

\bibitem{Sandin:2008db}
  F.~Sandin and P.~Ciarcelluti,
  Astropart.\ Phys.\  {\bf 32}, 278 (2009)
  [arXiv:0809.2942 [astro-ph]].
\bibitem{Ciarcelluti:2010ji}
  P.~Ciarcelluti, F.~Sandin,
  Phys.\ Lett.\  {\bf B695}, 19-21 (2011).
  [arXiv:1005.0857 [astro-ph.HE]].
\bibitem{Iorio:2010hb}
  L.~Iorio,
  JCAP {\bf 1011}, 046 (2010).
  [arXiv:1005.5078 [gr-qc]].

\bibitem{Goldman:1989nd}
  I.~Goldman and S.~Nussinov,
  Phys.\ Rev.\  D {\bf 40}, 3221 (1989).

\bibitem{Gould:1989gw}
  A.~Gould, B.~T.~Draine, R.~W.~Romani and S.~Nussinov,
  Phys.\ Lett.\  B {\bf 238}, 337 (1990).

\bibitem{Bertone:2007ae}
  G.~Bertone and M.~Fairbairn,
  Phys.\ Rev.\  D {\bf 77}, 043515 (2008)
  [arXiv:0709.1485 [astro-ph]].

\bibitem{Kouvaris:2010jy}
  C.~Kouvaris and P.~Tinyakov,
  arXiv:1012.2039 [astro-ph.HE].

\bibitem{Shapiro:1983du}
  S.~L.~Shapiro, S.~A.~Teukolsky,
  New York, USA: Wiley (1983) 645 p.

\bibitem{Clark:2002db}
  J.~S.~Clark, S.~P.~Goodwin, P.~A.~Crowther, L.~Kaper, M.~Fairbairn, N.~Langer and C.~Brocksopp,
  Astron.\ Astrophys.\  {\bf 392}, 909 (2002)
  [arXiv:astro-ph/0207334].



\bibitem{Zentner:2009is}
  A.~R.~Zentner,
  Phys.\ Rev.\  {\bf D80}, 063501 (2009).
  [arXiv:0907.3448 [astro-ph.HE]].


\bibitem{Feng:2009hw}
  J.~L.~Feng, M.~Kaplinghat and H.~B.~Yu,
  Phys.\ Rev.\ Lett.\  {\bf 104}, 151301 (2010)
  [arXiv:0911.0422 [hep-ph]].

\bibitem{Lin:2011gj}
  T.~Lin, H.~B.~Yu and K.~M.~Zurek,
  arXiv:1111.0293 [hep-ph].




\bibitem{Kepler:2006ns}
  S.~O.~Kepler {\it et al.},
  Mon.\ Not.\ Roy.\ Astron.\ Soc.\  {\bf 375}, 1315 (2007)
  [arXiv:astro-ph/0612277].





\bibitem{Kargaltsev:2003eb}
  O.~Kargaltsev, G.~G.~Pavlov and R.~W.~Romani,
  Astrophys.\ J.\  {\bf 602}, 327 (2004)
  [arXiv:astro-ph/0310854].


\bibitem{Manchester:2004bp}
  R.~N.~Manchester, G.~B.~Hobbs, A.~Teoh and M.~Hobbs,
  Astron.\ J.\  {\bf 129}, 1993 (2005)
  [arXiv:astro-ph/0412641].
  {\tt http://www.atnf.csiro.au/people/pulsar/psrcat/}

\bibitem{Gundmundsson}
 E.~H. Gudmundsson, C. J. Pethick and R. I. Epstein,
  Astrophys.\ J.\ {\bf 259}, L19 (1982).

\bibitem{globular}
{\tt http://www.naic.edu/\char`\~pfreire/GCpsr.html}

\bibitem{Mashchenko:2004hj}
 S.~Mashchenko, A.~Sills,
 Astrophys.\ J.\  {\bf 619}, 243 (2005).
 [astro-ph/0409605].
 S.~Mashchenko, A.~Sills,
 Astrophys.\ J.\  {\bf 619}, 258 (2005).
 [astro-ph/0409606].

\bibitem{Kouvaris:2011fi}
  C.~Kouvaris, P.~Tinyakov,
  [arXiv:1104.0382 [astro-ph.CO]].


\end{thebibliography}
\end{document}